\documentclass[11pt]{article}

\usepackage{mathpazo}
\usepackage[latin9]{inputenc}
\usepackage[a4paper]{geometry}
\usepackage{color}
\definecolor{note_fontcolor}{rgb}{0.800781, 0.800781, 0.800781}
\usepackage{units}
\usepackage{amsmath}
\usepackage{amssymb}
\usepackage{setspace}

\usepackage{comment}

\makeatletter

\@ifundefined{date}{}{\date{}}
\usepackage{graphicx}
\usepackage{calrsfs}
\usepackage[outerbars]{changebar}
\usepackage{soul}
\usepackage{microtype}

\DeclareMathOperator{\Tr}{Tr}

\makeatother

\global\long\def\kb#1#2{|#1\rangle\,\langle#2|}

\global\long\def\braket#1#2{\langle#1|#2\rangle}

\global\long\def\ket#1{|#1\rangle}

\global\long\def\bra#1{\langle#1|}

\begin{document}

	
	

	\title{Preparational Uncertainty Relations\\
		for $N$ Continuous Variables }
	
	\author{Spiros Kechrimparis\thanks{skechrimparis@gmail.com}\phantom{x}and Stefan Weigert\thanks{stefan.weigert@york.ac.uk}\\
		Department of Mathematics, University of York \\
		York, YO10 5DD, United Kingdom }

	\date{02 Oct 2016}
	
	\maketitle
	
	\begin{abstract}
		A smooth function of the second moments of $N$ continuous variables
		gives rise to an~uncertainty relation if it is bounded from below.
		We present a method to systematically derive such bounds by generalizing
		an approach applied previously to a single continuous variable. New~uncertainty relations are obtained for multi-partite systems that
		allow one to distinguish entangled from separable states. We also
		investigate the geometry of the ``uncertainty region'' in the $N(2N+1)$-dimensional
		space of moments. It is shown to be a convex set, and the points on its boundary are found to be in one-to-one
		correspondence with pure Gaussian states of minimal uncertainty. For
		a single degree of freedom, the boundary  can be visualized as one
		sheet of a ``Lorentz-invariant'' hyperboloid in the three-dimensional
		space of second~moments.
	\end{abstract}

\section{Introduction}

Uncertainty relations express limitations on the precision with which
one can measure specific properties of a quantum system, such as position
and momentum of a quantum particle. These~relations come in different
flavours. They may express the inability to \emph{prepare} a quantum
system in a state for which incompatible properties  possess exact
values. Alternatively, \emph{error-disturbance} uncertainty relations
refer to the constraints encountered when attempting to extract precise
values through measurements on a single system. Both cases point to
the uncertainty inherent in the quantum description of the world.

Heisenberg was the first to realize, in 1927, that  uncertainty relations
exist for quantum systems~\cite{heisenberg27}. His physical arguments
were quickly developed by Kennard \cite{kennard27}, Weyl \cite{weyl28},
Robertson \cite{robertson29} and Schr\"{o}dinger \cite{schroedinger30}.
Except for Heisenberg's paper, the focus of these contributions was
on preparational uncertainty, not yet clearly distinguished from measurement
uncertainty. In 1965,  Arthurs and Kelly presented a model of
joint measurement of position and momentum \cite{arthurs+65}, laying
the foundations for interest in error-disturbance uncertainty relations,
which has grown considerably over the last\textls[-20]{ two~de\-cades. Different
approaches rely on different concepts of error, which has led to lively
debates \cite{busch+13,ozawa03}.}

In recent years, the discussion of uncertainty relations has turned
from conceptual aspects to applications, in line with the overall
thrust of quantum information. For example, the first protocol of
quantum cryptography, known as BB84 \cite{bennett+84}, is based on pairs of mutual unbiased
bases that are known to come with maximal preparational uncertainty.
It is also possible to use variance-based uncertainty relations to
formulate criteria which detect entangled states of bi-partite systems \cite{duan+00,simon00}. 

This work investigates the structure of preparational uncertainty
relations in quantum systems with  more than one continuous variable,
i.e., $N\geq2$. Examples are given by a point particle moving in a
plane ($N=2$) or in three-dimensional space ($N=3)$; alternatively,
one may consider $N$ particles each moving along a real line,  each
with configuration space $\mathbb{R}$. Our main goals are (i) to
obtain lower bounds for given smooth functions depending on the $N(2N+1)$
second moments of a system with $N$ continuous variables, (ii) turn
these bounds into criteria that enable us to detect entangled states,
and (iii) to understand the geometric structure of uncertainty functionals
in the space of second moments, spanned by the independent elements
of the covariance matrix.

Using a variational technique originally introduced by Jackiw \cite{jackiw68},
we will generalize an approach that has been carried out successfully
for quantum systems with a single particle-type degree of freedom,
i.e., $N=1$ \cite{kechrimparis+15}. Encouraged by the new uncertainty
relations obtained  in this way for a single continuous variable,
we are particularly interested in the possibility to create inequalities
that are capable of detecting entangled states in systems with two or
more continuous variables. Tools to detect entanglement are crucial
for the implementation of any protocol in quantum information that
relies on entangled states. For continuous variables, quantum optical
methods are available to reliably check variance-based entanglement
criteria, allowing one to verify that a required entangled state has
indeed been created \cite{tasca+08,toscano+15,paul+16}. 

In Section  \ref{sec:Lower-bounds}, we will introduce uncertainty functionals
for $N$ continuous variables depending on second moments and describe
a method to determine their extrema and, subsequently, their minima.
Section \ref{sec:Inequalities} applies the approach to simple cases,
leading to new uncertainty relations, some of which may be used to
signal the presence of entangled states. A useful geometrical picture
of the uncertainty region---i.e., the covariance matrices
represented in the space of second moments---is derived
in Section~\ref{sec:Uncertainty-region}. The final section contains
a brief summary.

\section{Lower Bounds of Uncertainty Functionals \label{sec:Lower-bounds}}
\subsection{Extrema of Uncertainty Functionals \label{sec:Extrema} }

To describe a quantum system with $N$ continuous variables, one~associates $N$ pairs of canonical operators obeying the commutation relations 
\begin{align}
[\hat{q}_{k},\hat{p}_{k^{\prime}}]=i\hbar\delta_{kk^{\prime}}\,,\quad[\hat{q}_{k},\hat{q}_{k^{\prime}}]=[\hat{p}_{k},\hat{p}_{k^{\prime}}]=0\,,\quad k,k^{\prime}=1,\ldots,N\, 
\end{align}

We will arrange the momentum and position operators of the $k$-th
degree of freedom, $\hat{p}_{k}$ and $\hat{q}_{k}$, respectively,
into a column vector  $\mathbf{\hat{z}}$, 
\begin{align}
\mathbf{\hat{z}^{\top}}=(\hat{p}_{1},\hat{q}_{1},\ldots,\hat{p}_{N},\hat{q}_{N})\,\equiv(\hat{z}_{1},\hat{z}_{2},\ldots,\hat{z}_{2N-1},\hat{z}_{2N})\,
\end{align}
with components $\hat{z}_{\mu},\mu=1,\ldots,2N$. The pure states
of the quantum systems considered here are represented by unit vectors
$\ket{\psi}\in\mathcal{H}$, elements of an infinite-dimensional Hilbert space
$\mathcal{H}$. Of the $(2N)^{2}$ second~moments 
\begin{align}
c_{\mu\nu}=\frac{1}{2}\bra{\psi}\left(\hat{z}_{\mu}\hat{z}_{\nu}+\hat{z}_{\mu}\hat{z}_{\nu}\right)\ket{\psi}\,,\quad\mu\,,\nu=1,\ldots,2N\, \label{second moments definition}
\end{align}
only $N(2N+1)$ are independent. We assume (without loss of generality)
 that all first moments vanish, which follows from the invariance
of the second moments  under rigid phase-space translations. The~second
moments $c_{\mu\nu}$ form the \emph{covariance matrix} $\mathbf{C}$
associated with the pure state $\ket{\psi}$. 

With $k=1,\ldots,N,$ and for $\mu=\nu=2k-1$ ($\mu=\nu=2k$), we
obtain the variance of momentum (position) of the $k$-th degree
of freedom, while for $\mu=2k,\nu=2k-1$, we obtain their covariance;
all~other values of the indices $\mu,\nu$, correspond to moments
that mix different degrees of freedom.  Occasionally, we will denote
the variances of the $k$-th momentum and position with $x_{k}$ and
$y_{k}$, respectively, and their covariance by $w_{k}$.

Given a real function of the second moments for $N$ continuous variables,
 $f:\mathbb{R}^{N(2N+1)}$ $\rightarrow\mathbb{R}$,  we~wish to establish
whether it has a non-trivial lower bound $b$.  If it does, the statement
$f\geq b$ provides  an uncertainty relation. 

 Following an idea of Jackiw \cite{jackiw68} (see also \cite{busch+87,bialnicki+12,rudnicki12}), we define \emph{ }an\emph{
uncertainty functional} associated with the function $f$  by 
\begin{align}
J[\psi] & =f\left(\Delta^{2}p_{1},\Delta^{2}q_{1},C_{p_{1}q_{1}},\ldots C_{p_{1}p_{2}},C_{p_{1}q_{2}},\dots\right)-\lambda(\braket{\psi}{\psi}-1)\,\nonumber \\
 & =f\left(x_{1},y_{1},w_{1},\ldots,c_{13},c_{14},\dots\right)-\lambda(\braket{\psi}{\psi}-1)\, \label{UFN}
\end{align}
where the Lagrange multiplier $\lambda$ ensures that  any solutions
will be given by a  normalised state. We~first list  all \emph{local}
second moments for each degree of freedom (the two variances and the
covariance), followed by the \emph{non-local} moments which involve
different degrees of freedom. A variation of such a~functional will,
in analogy to the one-dimensional case (cf. \cite{kechrimparis+15,weigert96}), lead to an eigenvalue equation quadratic in position and momentum
operators. Let us briefly  spell out the derivation in the more general
setting.

First, we  compare the value of the functional $J[\psi]$ in the state
$\ket{\psi+\varepsilon}=\ket{\psi}+\varepsilon\ket e$ with its value
in the state $\ket{\psi}$, where $\ket e\in{\cal H}$ is an arbitrary
normalised state. Expanding it up to a second order in the small parameter
$\varepsilon$,  we find 
\begin{align}
J[\psi+\varepsilon]=J[\psi]+\varepsilon D_{\varepsilon}J[\psi]+O\left(\varepsilon^{2}\right) \label{FunctionalExpansion}
\end{align}
where the expression 
\begin{equation}
D_{\varepsilon}=\bra e\frac{\delta}{\delta\bra{\psi}}+\frac{\delta}{\delta\ket{\psi}}\ket e 
\end{equation}
denotes a G\^{a}teaux derivative.  The stationary points of the functional
are characterised by the vanishing of the first-order term in the
expansion (\ref{FunctionalExpansion}), 
\begin{align}
D_{\varepsilon}J[\psi]=\bra e\left(\frac{\delta}{\delta\bra{\psi}}f\left(x_{1},y_{1},w_{1},\ldots,c_{13},c_{14},\dots\right)-\lambda\ket{\psi}\right)+\text{c.c.}=0\, 
\end{align}

More explicitly, this condition reads 
\begin{align}
\bra e\left(\sum_{\mu\leq\nu}\left(\frac{\partial f}{\partial c_{\mu\nu}}\frac{\delta c_{\mu\nu}}{\delta\bra{\psi}}\right)-\lambda\ket{\psi}\right)+\text{c.c.}=0\,\label{variation}
\end{align}
 where the sum runs over the values $1\leq\mu\leq2N$ and $\mu\leq\nu\leq2N$. Since Equation \eqref{variation}
should hold for arbitrary variations of the ket $\ket e$ and its
dual $\bra e$ (which are independent), the expression in round brackets
as well as its complex conjugate must vanish identically. 

The functional derivatives of the second moments are 
\begin{equation}
\frac{\delta c_{\mu\nu}}{\delta\bra{\psi}}\equiv\frac{1}{2}\left(\hat{z}_{\mu}\hat{z}_{\nu}+\hat{z}_{\nu}\hat{z}_{\mu}\right)\ket{\psi}\, 
\end{equation}
resulting in a \emph{Euler-Lagrange}-type equation 
\begin{equation}
\left(\sum_{\mu\leq\nu}\frac{1}{2}\left(\hat{z}_{\mu}\hat{z}_{\nu}+\hat{z}_{\nu}\hat{z}_{\mu}\right)\frac{\partial f}{\partial c_{\mu\nu}}-\lambda\right)\ket{\psi}=0\, \label{eq: condition with lambda}
\end{equation}

The value of the multiplier $\lambda$ can be found  by multiplying
this equation with the bra $\bra{\psi}$ from the left and solving
for $\lambda$. Substituting its value back into Equation~(\ref{eq: condition with lambda}),
one finds the nonlinear eigenvector-eigenvalue equation 
\begin{equation}
\sum_{\mu\leq\nu}\frac{1}{2}\left(\hat{z}_{\mu}\hat{z}_{\nu}+\hat{z}_{\nu}\hat{z}_{\mu}\right)\frac{\partial f}{\partial c_{\mu\nu}}\ket{\psi}=\sum_{\mu\leq\nu}c_{\mu\nu}\frac{\partial f}{\partial c_{\mu\nu}}\ket{\psi}\, \label{EigEq}
\end{equation}
or, in matrix notation, 
\begin{equation}
\left(\hat{\mathbf{z}}^{\top}\mathbf{F}\ \hat{\mathbf{z}}\right)\ket{\psi}=\Tr\left(\mathbf{C\,}\mathbf{F}\right)\ket{\psi}\,\label{EigenEqMatr}
\end{equation}
where the matrix $\mathbf{F}$ is defined in terms of the first partial
derivatives of the function $f$: its diagonal elements are equal
to $f_{c_{\mu\mu}}$, while the off-diagonal ones are given by $f_{c_{\mu\nu}}/2$ with $\mu\neq\nu$,
using the standard convention to denote partial derivatives by subscripts.
As an example, the eigenvalue equation becomes, for $N=2$,
\begin{eqnarray}
\left(\sum_{k=1}^{2}\left(f_{x_{k}}\hat{p}_{k}^{2}+f_{y_{k}}\hat{q}_{k}^{2}+\frac{f_{w_{k}}}{2}\left(\hat{q}_{k}\hat{p}_{k}+\hat{p}_{k}\hat{q}_{k}\right)\right)+f_{c_{13}}\hat{p}_{1}\hat{p}_{2}+\ldots+f_{c_{24}}\hat{q}_{1}\hat{q}_{2}\right)\ket{\psi}=\nonumber \\
=\left(\sum_{k=1}^{2}\left(x_{k}f_{x_{k}}+y_{k}f_{y_{k}}+w_{k}f_{z_{k}}\right)+c_{13}f_{c_{13}}+\ldots+c_{24}f_{c_{24}}\right)\ket{\psi}\,\nonumber \\
\label{EigCor}
\end{eqnarray}

Note that Equation \eqref{EigenEqMatr} is generally non-linear in the
state $\ket{\psi}$ since the second moments and the partial derivatives
of $f$ are functions of expectation values in the state $\ket{\psi}$.
As we will show in next section, one can nevertheless solve Equation \eqref{EigenEqMatr},
given a number of assumptions.

\subsection{Consistency Conditions \label{sec:Consistency}}

To solve Equation \eqref{EigenEqMatr}, we initially assume that the matrix
$\mathbf{F}$ of partial derivatives is \emph{constant}, i.e., we suppress
its dependence on the state $\ket{\psi}$. If we further require that
$\mathbf{F}$ is positive definite, then~Williamson's theorem \cite{williamson36,simon+94}
guarantees the existence of a symplectic matrix $\mathbf{\Sigma}$
that puts $\mathbf{F}$ into a~diagonal form, i.e., 
\begin{equation}
\mathbf{F}=\mathbf{\Sigma}^{\top}\mathbf{D}\,\mathbf{\Sigma}\,
\end{equation}
where the diagonal matrix $\mathbf{D}$ is defined by $\mathbf{D}=\text{diag}(\lambda_{1},\lambda_{1},\ldots,\lambda_{N},\lambda_{N})$,
and the positive real numbers $\lambda_{k}>0$, $k=1,\ldots,N$, are
the \emph{symplectic eigenvalues} of $\mathbf{F}$ \cite{dutta+95,adesso+14,simon+94}.
We recall that a symplectic matrix of order $2N$ satisfies $\mathbf{\Sigma}^{\top}\mathbf{\Omega}\mathbf{\,\Sigma}=\mathbf{\Omega}$,
where $\mathbf{\Omega}$ is uniquely determined by the commutation
relations, $[\hat{z}_{\mu},\hat{z}_{\nu}]=i\hbar\Omega_{\mu\nu}$,
$\mu,\nu=1,\ldots,2N$.

Multiplying both sides of Equation \eqref{EigenEqMatr} with the metaplectic
unitary operator $\hat{S}^{\dagger}$ from the left, defined by the
relation
\begin{equation}
\mathbf{\Sigma}\ \hat{\mathbf{z}}=\hat{S}\hat{\,\mathbf{z}}\,\hat{S}^{\dagger}\, \label{metapl}
\end{equation}
we find that its left-hand-side can be expressed as 
\begin{align}
\hat{S}^{\dagger}\left(\hat{\mathbf{z}}^{\top}\mathbf{F}\ \hat{\mathbf{z}}\right)\hat{S}\left(\hat{S}^{\dagger}\ket{\psi}\right) & =\left(\hat{S}^{\dagger}\hat{\mathbf{z}}^{\top}\hat{S}\right)\mathbf{F}\left(\hat{S}^{\dagger}\hat{\mathbf{z}}\hat{S}\right)\left(\hat{S}^{\dagger}\ket{\psi}\right)\nonumber \\
 & =\left(\mathbf{\Sigma}^{-1}\hat{\mathbf{z}}\right)^{\top}\left(\mathbf{\Sigma}^{\top}\mathbf{D}\,\mathbf{\Sigma}\right)\left(\mathbf{\Sigma}^{-1}\hat{\mathbf{z}}\right)\left(\hat{S}^{\dagger}\ket{\psi}\right)\,
\end{align}

Thus, Equation \eqref{EigenEqMatr}   simplifies to 
\begin{align}
\left(\hat{\mathbf{z}}^{\top}\mathbf{D}\ \hat{\mathbf{z}}\right)\left(\hat{S}^{\dagger}\ket{\psi}\right)=\Tr\left(\mathbf{C}\,\mathbf{F}\right)\left(\hat{S}^{\dagger}\ket{\psi}\right)\, 
\end{align}
which can be written as 
\begin{align}
\sum_{k=1}^{N}\lambda_{k}\left(\frac{\hat{p}_{k}^{2}+\hat{q}_{k}^{2}}{2}\right)\left(\hat{S}^{\dagger}\ket{\psi}\right)=\frac{1}{2}\Tr\left(\mathbf{C}\,\mathbf{F}\right)\left(\hat{S}^{\dagger}\ket{\psi}\right)\,\label{eq: N decoupled oscillators}
\end{align}

Thus, we have transformed the quadratic operator on the left-hand-side
of Equation \eqref{EigenEqMatr} into a~Hamiltonian operator given by a
sum of  $N$ decoupled harmonic oscillators. The solutions of Equation~\eqref{eq: N decoupled oscillators} are given by tensor products
of number states for each degree of freedom: 
\begin{equation}
\ket{\psi}=\hat{S}\left(\ket{n_{1}}\otimes\ldots\otimes\ket{n_{N}}\right)\equiv\hat{S}\left(\underset{k=1}{\bigotimes^{N}}\ket{n_{k}}\right) \label{ExtremaNDofs}
\end{equation}

Note that the constraint 
\begin{equation}
\frac{1}{2}\Tr\left(\mathbf{C}\,\mathbf{F}\right)=\sum_{k=1}^{N}\lambda_{k}\left(n_{k}+\frac{1}{2}\right)\hbar\, \label{linCC}
\end{equation}
must be satisfied by all potential extremal states.

Recall that we have treated the matrix elements of the matrix $\mathbf{F}$
introduced in Equation \eqref{EigenEqMatr}  as constants, on which
the  unitary transformation $\hat{S}$ and hence the states $\ket{\psi}$
in Equation \eqref{ExtremaNDofs} now depend. To achieve consistency, we
determine the expectation value of the covariance matrix in the solution~$\ket{\psi}$.  A set of coupled equations in matrix form results
for the extremal second moments, which we will call the \emph{consistency
conditions}. Explicitly, we find
\begin{align}
\mathbf{C} & =\bra{\psi}\mathbf{\hat{C}}\ket{\psi}\equiv\frac{1}{2}\bra{\psi}\left(\hat{\mathbf{z}}\otimes\hat{\mathbf{z}}^{\top}+\left(\hat{\mathbf{z}}\otimes\hat{\mathbf{z}}^{\top}\right)^{\top}\right)\ket{\psi}\nonumber \\
 & =\frac{1}{2}\left(\bigotimes_{k=1}^{N}\bra{n_{k}}\right)\hat{S}^{\dagger}\left(\hat{\mathbf{z}}\otimes\hat{\mathbf{z}}^{\top}+\left(\hat{\mathbf{z}}\otimes\hat{\mathbf{z}}^{\top}\right)^{\top}\right)\hat{S}\left(\bigotimes_{k^{\prime}=1}^{N}\ket{n_{k^{\prime}}}\right)\,
\end{align}
where $\hat{\mathbf{z}}\otimes\hat{\mathbf{z}}^{\top}$ denotes the
Kronecker product of the column vector $\mathbf{\hat{z}}$ with its
transpose, $\hat{\mathbf{z}}^{\top}$. Using the identity \eqref{metapl}
in the form $\mathbf{\Sigma^{-1}}\,\hat{\mathbf{z}}=\hat{S}^{\dagger}\hat{\,\mathbf{z}}\,\hat{S}$,
we can express the covariance matrix in the form 

\begin{equation}
\mathbf{C}=\mathbf{\Sigma}^{-1}\frac{1}{2}\left(\mathbf{N}+\mathbf{N}^{\top}\right)(\mathbf{\Sigma}^{-1})^{\top}\, 
\end{equation}
with the matrix 
\begin{equation}
\mathbf{N}=\left(\bigotimes_{k=1}^{N}\bra{n_{k}}\right)\hat{\mathbf{z}}\otimes\hat{\mathbf{z}}^{\top}\left(\bigotimes_{k^{\prime}=1}^{N}\ket{n_{k^{\prime}}}\right)\, 
\end{equation}
having elements
\begin{equation}
N_{\mu\nu}=\bra{n_{1},\ldots,n_{N}}\,\hat{z}_{\mu}\hat{z}_{\nu}\,\ket{n_{1},\ldots,n_{N}}\,,\quad\mu,\nu=1,\ldots,2N\,
\end{equation}

Recalling that the components of the vector $\hat{\mathbf{z}}$ are
position and momentum operators, it is not difficult to see that the
only non-zero matrix elements of $\mathbf{N}$ are on its diagonal,
i.e., 
\begin{equation}
\mathbf{N}=\hbar\text{diag}\left(n_{1}+\frac{1}{2},n_{1}+\frac{1}{2},\ldots,n_{N}+\frac{1}{2},n_{N}+\frac{1}{2}\right)
\end{equation}

Using the property $\mathbf{N}^{\top}=\mathbf{N}$, which holds for
any diagonal matrix, we finally obtain the \emph{consistency conditions}
for $N$ continuous variables,
\begin{equation}
\mathbf{C}=\mathbf{\Sigma}^{-1}\mathbf{N}\,(\mathbf{\Sigma}^{-1})^{\top}\, \label{CCs}
\end{equation}

These conditions select the  extrema that are compatible with the
specific function of the second~moments considered. The constraint
given in \eqref{linCC} can be rewritten as 
\begin{align}
\Tr(\mathbf{C}\,\mathbf{F})=\Tr(\mathbf{D}\,\mathbf{N})\, 
\end{align}
and it is easy to check that this condition is trivially satisfied
if the consistency conditions \eqref{CCs} hold.

The take-away message from the conditions \eqref{CCs} can be summarised
as follows: \emph{a function $f$ of the second moments of $N$ positions
and momenta has an extremum in a pure state $\ket{\psi}$ if there
exists a symplectic matrix }$\mathbf{\Sigma}$\emph{ that diagonalises
the  covariance matrix $\mathbf{C}$ and, at the same time, the transpose
of its inverse, $\left(\mathbf{\Sigma}^{-1}\right)^{\top}$, diagonalises
the matrix $\mathbf{F}$ of the partial derivatives of the function
$f$}.

According to \eqref{CCs}, the determinant of the covariance matrix
 for extremal states of the uncertainty functional $J[\psi]$ takes
the value 
\begin{equation}
\det\mathbf{C}=\prod_{k=1}^{N}\left(n_{k}+\frac{1}{2}\right)^{2}\hbar^{2}\, \label{eq: det for extremal states}
\end{equation}

Clearly, the minimum is achieved when each oscillator resides in its
ground state, 
\begin{equation}
\det\mathbf{C}\geq\left(\frac{\hbar}{2}\right)^{2N}\,\label{eq: det form of RS for N particles}
\end{equation}
corresponding to $n_{1}=\ldots=n_{N}=0$ in Equation \eqref{eq: det for extremal states}.

No pure $N$-particle state can give rise to a covariance matrix $\mathbf{C}$
violating the inequality \eqref{eq: det form of RS for N particles}.
This universally valid constraint generalizes the single-particle
inequality derived by Robertson and Schr\"{o}dinger to $N$ particles,
expressing it elegantly as a condition on the determinant of the covariance
matrix of a state. Supplying \eqref{eq: det for extremal states}
with the lower-dimensional Robertson-Schr\"{o}dinger-type inequalities
that need to be obeyed in by each subsystem of dimension $2$ to $N-1$,
we get the general uncertainty statement for more than one degrees
of freedom, usually expressed in the form, 
\begin{equation}
\mathbf{C}+i\frac{\hbar}{2}\mathbf{\Omega}\geq0\,
\end{equation}

Alternatively, this requirement can be expressed in terms of  inequalities
for the symplectic eigenvalues of the covariance matrix \cite{simon+94,adesso+14}.

We conclude this section by explicitly working out the consistency
conditions for one degree of freedom, $N=1$. In this case, we obtain
the matrices  $\mathbf{N}=\hbar(n+1/2)\mathbf{I}$ and $\mathbf{\Sigma}=\mathbf{S}_{\gamma}\mathbf{G}_{b}$,
with  symplectic matrices $\mathbf{G}_{b}$ and $\mathbf{S}_{\gamma}$
given by 
\begin{equation}
\textbf{G}_{b}=\left(\begin{array}{cc}
1 & 0\\
b & 1
\end{array}\right)\,,\mbox{\quad and}\quad\mathbf{S}_{\gamma}=\left(\begin{array}{cc}
e^{-\gamma} & 0\\
0 & e^{\gamma}
\end{array}\right)\, \label{eq: define G and S}
\end{equation}
respectively, and real parameters 
\begin{equation}
b=\frac{f_{w}}{2f_{y}}\in\mathbb{R}\quad\mbox{and}\quad\gamma=\frac{1}{2}\ln\left(\frac{f_{y}}{\sqrt{\text{det}\,\textbf{F}}}\right)\,\in\mathbb{R}\, 
\end{equation}

The consistency conditions now take the simple form 
\begin{align}
\mathbf{C} & =\mathbf{\Sigma}^{-1}\mathbf{N}(\mathbf{\Sigma}^{-1})^{\top}=\mathbf{G}_{b}^{-1}\mathbf{S}_{\gamma}^{-1}(\mathbf{S}_{\gamma}^{-1})^{\top}(\mathbf{G}_{b}^{-1})^{\top}\hbar\left(n+\frac{1}{2}\right)=\mathbf{F}^{-1}\hbar\left(n+\frac{1}{2}\right)\,\sqrt{\text{det}\,\textbf{F}}
\end{align}
or finally, 
\begin{equation}
\frac{\text{{\bf F}}\ \text{{\bf C}}}{\sqrt{\text{det}\,\textbf{F}}}=\hbar\left(n+\frac{1}{2}\right)\,\mathbf{I}\,,\qquad n\in\mathbb{N}_{0}\, \label{eq: consistency N=00003D00003D1}
\end{equation}

Therefore, the formalism developed here correctly reproduces the findings
of \cite{kechrimparis+15}.

\section{Inequalities for Two or More Continuous Variables\label{sec:Inequalities}}
\subsection{Inequalities without Correlation Terms \label{sec:NoCorrelations}}

Let us now examine the consistency conditions for more than one degree
of freedom while allowing only \emph{product} states. Correlations between
the degrees of freedom being absent, the functional will only depend
on the \emph{local }second moments, i.e., $f\equiv f(x_{1},y_{1},w_{1},\ldots,x_{N},y_{N},w_{N})$;
the $2N(N-1)$ moments mixing the degrees of freedom are always zero
in a separable state. For simplicity, we only consider $N=2$ in some
detail, the generalisation to $N>2$ being straightforward.

Using matrices $\mathbf{G}_{b}$ and $\mathbf{S}_{\gamma}$ defined
in \eqref{eq: define G and S}, we construct two symplectic matrices
$\mathbf{S}_{1}$ and $\mathbf{S}_{2}$ as~follows: 
\begin{equation}
\mathbf{\Sigma}_{1}=\left(\begin{array}{cc}
\mathbf{S}_{\gamma_{1}}\textbf{G}_{b_{1}} & 0\\
0 & \mathbf{I}
\end{array}\right)\quad\mbox{and}\quad\mathbf{\Sigma}_{2}=\left(\begin{array}{cc}
\mathbf{I} & 0\\
0 & \mathbf{S}_{\gamma_{2}}\textbf{G}_{b_{2}}
\end{array}\right)\,
\end{equation}

Their product, $\mathbf{\Sigma}=\mathbf{\Sigma}_{1}\mathbf{\Sigma}_{2},$
describes the action of the factorised unitary operator 
\begin{equation}
\hat{S}=\hat{S}_{1}\otimes\hat{S}_{2}\, 
\end{equation}
when solving the eigenvalue Equation \eqref{EigenEqMatr}. The consistency
conditions become 
\begin{align}
\mathbf{C} & =\mathbf{\Sigma}^{-1}\mathbf{N}(\mathbf{\Sigma}^{-1})^{\top}=\mathbf{\Sigma}^{-1}(\mathbf{\Sigma}^{-1})^{\top}\mathbf{N}=\mathbf{F}_{pr}^{-1}\mathbf{N}\,
\end{align}
with
\begin{align}
\mathbf{F}_{pr}=\left(\begin{array}{cc}
\mathbf{F}_{1}/\sqrt{\text{det}\,\textbf{F}_{1}} & 0\\
0 & \mathbf{F}_{2}/\sqrt{\text{det}\,\textbf{F}_{2}}
\end{array}\right)\, \label{Fpr}
\end{align}
 so that we finally obtain 
\begin{align}
\mathbf{F}_{pr}\mathbf{C}=\mathbf{N}\, \label{CCsProd}
\end{align}

In Equation \eqref{Fpr}, the $2\times2$ matrices $\mathbf{F}_{k},k=1,2$,
denote the collection of partial derivatives of the function $f$
with respect to the moments of the $k$-th degree of freedom. Therefore,
 the consistency conditions for functionals of product states reduce
to a pair of one-dimensional ones that must be solved simultaneously.

The generalisation to $N$ degrees of freedom is straightforward:
   for each extra degree of freedom, a matrix $\mathbf{F}_{k}/\sqrt{\text{det}\,\textbf{F}_{k}}$
must be  added to the diagonal of the block matrix $\mathbf{F}_{pr}$.
After introducing the suitably generalized matrices $\mathbf{C}$
and $\mathbf{N}$, Equation \eqref{CCsProd}  describes the consistency
conditions for  separable quantum states. It is often useful to express
Equation \eqref{CCsProd} as 
\begin{align}
x_{k}f_{x_{k}}=y_{k}f_{y_{k}}\,,\quad2w_{k}f_{y_{k}}=-x_{k}f_{w_{k}}\,,\quad x_{k}y_{k}-w_{k}^{2}=\hbar^{2}\left(n_{k}+\frac{1}{2}\right)\,
\end{align}
with $k=1,\ldots,N$.

The simplest example of a factorized uncertainty relation is given
by the product of  two one-dimensional Robertson-Schr\"{o}dinger inequalities,
 following from the functional 
\begin{equation}
f(x_{1},y_{1},w_{1},x_{2},y_{2},w_{2})=(x_{1}y_{1}-w_{1}^{2})(x_{2}y_{2}-w_{2}^{2})\,
\end{equation}

The resulting inequality, 
\begin{equation}
\left(\Delta^{2}p_{1}\,\Delta^{2}q_{1}-C_{p_{1}q_{1}}^{2}\right)\left(\Delta^{2}p_{2}\,\Delta^{2}q_{2}-C_{p_{2}q_{2}}^{2}\right)\geq\left(\frac{\hbar}{2}\right)^{4}\, \label{RobDOF}
\end{equation}
corresponds to the boundary described by Equation \eqref{eq: det form of RS for N particles}
in the absence of correlations, to~be discussed in more detail in
Section \ref{sec:Uncertainty-region}. Note that this inequality is only
invariant under $\mbox{Sp}(2,\mathbb{R})\otimes\mbox{Sp}(2,\mathbb{R})$
transformations instead of those of the  $\mbox{Sp}(4,\mathbb{R})$
group that leave invariant the Robertson--Schr\"{o}dinger-type inequality
for two degrees of freedom.  However, the matrix inequality $\mathbf{C}+i\mathbf{\Omega}\hbar/2\geq0$
is invariant under any symplectic transformation and serves as the
required~generalisation.

 Starting from the functional 
\begin{equation}
f(x_{1},y_{1},w_{1},x_{2},y_{2},w_{2})=x_{1}\,y_{1}x_{2}\,y_{2}-w_{1}^{2}\,w_{2}^{2}
\end{equation}
we arrive -after solving \eqref{CCsProd}- at
\begin{equation}
\Delta^{2}p_{1}\,\Delta^{2}q_{1}\Delta^{2}p_{2}\,\Delta^{2}q_{2}\geq\left(\frac{\hbar}{2}\right)^{4}+C_{p_{1}q_{1}}^{2}C_{p_{2}q_{2}}^{2}\, 
\end{equation}
which cannot be obtained by a combination of inequalities for $N=1$.
It is \emph{stronger} than the (factorized) ``Heisenberg''-type
inequality for more than two observables 
\begin{equation}
\Delta p_{1}\,\Delta q_{1}\,\Delta p_{2}\,\Delta q_{2}\geq\left(\frac{\hbar}{2}\right)^{2}\, 
\end{equation}
first mentioned in a paper by Robertson \cite{robertson34}, but \emph{weaker}
than \eqref{RobDOF}.  An inequality $I_{1}$ is said to be \emph{weaker}
than the inequality $I_{2}$ if \emph{fewer} states saturate $I_{1}$ than
$I_{2}$.

Mixing products of variances related to different degrees of freedom
also leads to non-trivial inequalities such as  
\begin{align}
a\left(\Delta^{2}p_{1}\Delta^{2}q_{2}\right)^{n}+b\left(\Delta^{2}p_{2}\Delta^{2}q_{1}\right)^{n}\geq2\sqrt{ab}\left(\frac{\hbar}{2}\right)^{2n}\,,\quad a,b>0\, 
\end{align}

 For $a=b=1$ and $n=1$, one obtains 
\begin{align*}
\Delta p_{1}\Delta q_{2}+\Delta p_{2}\Delta q_{1} & \geq\hbar\, 
\end{align*}
which resembles the inequality for the sum of two one-dimensional
Heisenberg inequalities,  
\begin{equation}
\Delta p_{1}\Delta q_{1}+\Delta p_{2}\Delta q_{2}\geq\hbar\, 
\end{equation}
but differs fundamentally from it.

\subsection{Inequalities with Correlation Terms}

Dropping the limitation to product states, we now turn to  functionals
that involve terms to which  different degrees of freedom contribute.
To begin, let us consider  a linear combination of second~moments,
\begin{align*}
f\left(\Delta^{2}p_{1},\ldots,C_{q_{1}q_{2}}\right)=a\left(\Delta^{2}p_{1}+\Delta^{2}q_{1}\right)+b\left(\Delta^{2}p_{2}+\Delta^{2}q_{2}\right)+c\left(C_{p_{1}p_{2}}-C_{q_{1}q_{2}}\right)\, 
\end{align*}
for which the matrix $\mathbf{F}$ takes the form 
\begin{align}
\mathbf{F}=\left(\begin{array}{cccc}
a & 0 & \nicefrac{c}{2} & 0\\
0 & a & 0 & -\nicefrac{c}{2}\\
\nicefrac{c}{2} & 0 & b & 0\\
0 & -\nicefrac{c}{2} & 0 & b
\end{array}\right)\, 
\end{align}

It is positive definite whenever the coefficients $a,b,c$ obey the
conditions $a,b>0$ and $4ab>c^{2}$, which we assume from now on.
The symplectic matrix $\mathbf{S}$ that brings $\mathbf{F}$ to diagonal
form is given by (cf.~\cite{weedbrook+12}): 
\begin{align}
\mathbf{\Sigma}=\left(\begin{array}{cccc}
\sigma_{+} & 0 & \sigma_{-} & 0\\
0 & \sigma_{+} & 0 & -\sigma_{-}\\
\sigma_{-} & 0 & \sigma_{+} & 0\\
0 & -\sigma_{-} & 0 & \sigma_{+}
\end{array}\right)\, 
\end{align}
where 
\begin{equation}
\sigma_{\pm}=\sqrt{\frac{a+b\pm\sqrt{y}}{2\sqrt{y}}}\,,\quad\text{and}\quad y=(a+b)^{2}-c^{2}\, 
\end{equation}

The consistency conditions \eqref{CCs} can be solved in closed form,
 leading to the covariance matrix at \mbox{the extrema} 
\begin{align}
\mathbf{C} & =\left(\begin{array}{cccc}
\Delta^{2}p_{1}^{(e)} & 0 & C_{p_{1}p_{2}}^{(e)} & 0\\
0 & \Delta^{2}q_{1}^{(e)} & 0 & C_{q_{1}q_{2}}^{(e)}\\
C_{p_{1}p_{2}}^{(e)} & 0 & \Delta^{2}p_{2}^{(e)} & 0\\
0 & C_{q_{1}q_{2}}^{(e)} & 0 & \Delta^{2}q_{2}^{(e)}
\end{array}\right)\, 
\end{align}
with elements explicitly given by 
\begin{align}
\Delta^{2}p_{1}^{(e)} & =\Delta^{2}q_{1}^{(e)}=\frac{(n_{1}-n_{2})\hbar}{2}+\frac{(a+b)(n_{1}+n_{2}+1)\hbar}{2\sqrt{(a+b)^{2}-c^{2}}}\, \label{eq:extremalonevariances}\\
\Delta^{2}p_{2}^{(e)} & =\Delta^{2}q_{2}^{(e)}=\frac{(n_{2}-n_{1})\hbar}{2}+\frac{(a+b)(n_{1}+n_{2}+1)\hbar}{2\sqrt{(a+b)^{2}-c^{2}}}\, \label{eq:extremaltwovariances}
\end{align}
and 
\begin{equation}
C_{p_{1}p_{2}}^{(e)}=-C_{q_{1}q_{2}}^{(e)}=-\frac{c(n_{1}+n_{2}+1)\hbar}{2\sqrt{(a+b)^{2}-c^{2}}}\, 
\end{equation}

One can check that the expressions on the right-hand side of Equations
\eqref{eq:extremalonevariances} and \eqref{eq:extremaltwovariances}
are \linebreak  positive, while
\begin{equation}
\left(C_{p_{1}p_{2}}^{(e)}\right)^{2}\leq\Delta^{2}p_{1}^{(e)}\,\Delta^{2}p_{2}^{(e)}\quad\mbox{and}\quad\left(C_{q_{1}q_{2}}^{(e)}\right)^{2}\leq\Delta^{2}q_{1}^{(e)}\,\Delta^{2}q_{2}^{(e)}
\end{equation}
also hold, as required. In fact, these two inequalities are never
saturated by the extremal states, although one can get arbitrarily
close if $n_{1}$ is zero, while $n_{2}$ tends to infinity (or {vice
versa).}

Substituting the extremal values of the second moments back into the
functional, we find 
\begin{align}
f_{a,b,c}^{(e)}(n_{1},n_{2})=(a-b)(n_{1}-n_{2})\hbar+\sqrt{(a+b)^{2}-c^{2}}(n_{1}+n_{2}+1)\hbar\geq f_{a,b,c}(0,0)
\end{align}
implying the following inequality, satisfied by any quantum state:
\begin{align}
a\left(\Delta^{2}p_{1}+\Delta^{2}q_{1}\right)+b\left(\Delta^{2}p_{2}+\Delta^{2}q_{2}\right)+c\left(C_{p_{1}p_{2}}-C_{q_{1}q_{2}}\right)\geq\hbar\sqrt{(a+b)^{2}-c^{2}}\, \label{CorIneq}
\end{align}

Pure separable states are known to satisfy the relation 
\begin{align}
a\left(\Delta^{2}p_{1}+\Delta^{2}q_{1}\right)+b\left(\Delta^{2}p_{2}+\Delta^{2}q_{2}\right)\geq(a+b)\hbar\,\label{UncorIneq}
\end{align}

Now consider the limit $c\rightarrow\pm2\sqrt{ab}$ in \eqref{CorIneq}
which, however, breaks the positive definiteness of $\mathbf{F}$: its right-hand-side
tends to zero and the terms on the left are just the sum of the variances
of the Einstein-Podolsky-Rosen-type (EPR) operators $\hat{u}_{1}=\sqrt{a}\,\hat{p}_{1}+\sqrt{b}\,\hat{p}_{2}$
and $\hat{u}_{2}=\sqrt{a}\,\hat{q}_{1}-\sqrt{b}\,\hat{q}_{2}$ \cite{duan+00,simon00}.
In this case, the~pair of inequalities \eqref{CorIneq} and \eqref{UncorIneq}
form the prototypical example of using uncertainty relations for entanglement~detection. More specifically, whenever the sum of the variances of
$\hat{u}_{1}$ and $\hat{u}_{2}$ in a given state $\ket{\psi}$ violates
the bound of \eqref{UncorIneq}, then the state is entangled. Although
inequality \eqref{UncorIneq} provides only a sufficient condition
for inseparability of an arbitrary state, it can become a sufficient
and necessary condition for pure Gaussian states, if recast in an
appropriate form \cite{duan+00}.

Returning to inequality \eqref{CorIneq} in the case of arbitrary
$a,b,c$, it is not immediately obvious whether it can be used to
detect entangled states. However, let us define four EPR-type operators:
\begin{align}
\hat{u}_{1} & =\alpha_{1}\hat{p}_{1}+\beta_{1}\hat{p}_{2}\,,\quad\hat{v}_{1}=\gamma_{1}\hat{q}_{1}-\delta_{1}\hat{q}_{2}\nonumber \\
\hat{u}_{2} & =\alpha_{2}\hat{p}_{1}+\beta_{2}\hat{p}_{2}\,,\quad\hat{v}_{2}=\gamma_{2}\hat{q}_{1}-\delta_{2}\hat{q}_{2}\,
\end{align}
with eight real parameters $\alpha_{1},\ldots, \delta_{2}$, which are
constrained by the relations 
\begin{align}
 & \alpha_{1}^{2}+\alpha_{2}^{2}=\gamma_{1}^{2}+\gamma_{2}^{2}=a\,,\quad\beta_{1}^{2}+\beta_{2}^{2}=\delta_{1}^{2}+\delta_{2}^{2}=b\,\nonumber \\
 & \qquad\quad\alpha_{1}\beta_{1}+\alpha_{2}\beta_{2}=\gamma_{1}\delta_{1}+\gamma_{2}\delta_{2}=c/2\,
\end{align}

Now, we can write Equation \eqref{CorIneq} as 
\begin{equation}
\Delta^{2}u_{1}+\Delta^{2}v_{1}+\Delta^{2}u_{2}+\Delta^{2}v_{2}\geq\hbar\sqrt{(a+b)^{2}-c^{2}}\, \label{CorFourEPR}
\end{equation}
reducing to the inequality 
\begin{align}
\Delta^{2}u_{1}+\Delta^{2}v_{1}+\Delta^{2}u_{2}+\Delta^{2}v_{2}\geq\hbar(a+b)\label{UncorFourEPR}
\end{align}
if the the system resides in a separable state. Since its right-hand-side
is always greater than or equal to the bound in \eqref{CorFourEPR},
the violation of \eqref{UncorFourEPR} indicates the presence of an
entangled state.

Clearly, inequality \eqref{CorFourEPR} is more general than the corresponding
one for the pair of operators $\hat{u}_{1}=\sqrt{a}\,\hat{p}_{1}+\sqrt{b}\,\hat{p}_{2}$
and $\hat{u}_{2}=\sqrt{a}\,\hat{q}_{1}-\sqrt{b}\,\hat{q}_{2}$, as
the former reduces to the latter in the limit $c\rightarrow\pm2\sqrt{ab}$
and thus extends a known result \cite{duan+00}.

As a final example, consider the sum of the variances of the EPR-type
operators for \emph{three} degrees of freedom, $\hat{u}_{1}=\hat{q}_{1}+\hat{p}_{2}+\hat{q}_{3}$,
$\hat{u}_{2}=\hat{q}_{2}+\hat{p}_{3}+\hat{q}_{1}$, $\hat{u}_{3}=\hat{q}_{3}+\hat{p}_{1}+\hat{q}_{2}$,
which is in general only bounded by zero. However, the lower possible
value achievable in a \emph{separable }state is given by the inequality
\begin{equation}
\Delta^{2}u_{1}+\Delta^{2}u_{2}+\Delta^{2}u_{3}\geq3\sqrt{2} \, \hbar\, \label{TripleSep}
\end{equation}
readily obtained from the solution of Equation \eqref{CCsProd}. Again,
violations of \eqref{TripleSep} detect the presence of entangled
degrees of freedom.

It is, of course, possible to minimise other functions than the sum
of the variances, leading to different entanglement-detecting inequalities
that we will discuss elsewhere. 

\section{The Uncertainty Region \label{sec:Uncertainty-region}}

In this section, we will develop a geometric view of quantum uncertainty
for a system with $N$ continuous variables. To do so, we associate
a direction of the space $\mathbb{R}^{d}$ with each of the second~moments $C_{\mu\nu},\mu,\nu=1,\ldots,2N$. Then, any quantum state
gives rise to a point in the \emph{space of second moments}, ${\cal S},$
which has dimension $d=N(2N+1)$.

Some points in the space ${\cal S}=\mathbb{R}^{d}$ will represent moments
of quantum states while others will~not. The accessible part of the
space is called the \emph{uncertainty region}, as the points it contains
are in one-to-one correspondence with admissible covariance matrices
$\mathbf{C}\in\mathbb{R}^{2N\times2N}$. This region is bounded by
a~$(d-1)$-dimensional surface given by the relation
\begin{equation}
\det\left(\mathbf{C}+i\frac{\hbar}{2}\mathbf{\Omega}\right)=0\, \label{eq: N-dim boundary}
\end{equation}
where $\mathbf{\Omega}$ is the standard symplectic matrix of order
$2N\times2N$.

\subsection{More Than One Continuous Variable: $N>1$\label{sub: More-than-one}}

We will show now that the uncertainty region in the space ${\cal S}$
is a \emph{convex }set, by affirming (i) that its \emph{boundary}
\eqref{eq: N-dim boundary} is convex and (ii) that all points of
the uncertainty region emerge as expectations taken in \emph{pure}
states. In other words, the uncertainty region has no ``pure-state
holes.'' This property justifies our initial decision to search for
extrema of uncertainty functionals among pure states only: no other
extrema would result had we included mixed states. On the boundary
of the uncertainty region, the relationship between quantum states and
their moments is unique (up to rigid translations) while (iii) points
inside the uncertainty region can also be obtained from infinitely
many different convex combinations of pure (or mixed) states.

\subsubsection{The Uncertainty Region Has a Convex Boundary}

The region defined by Equation \eqref{eq: det form of RS for N particles}
is a \emph{convex} set in the $N(2N+1)$-dimensional space of second
moments. To see this, we consider two covariance matrices $\mathbf{C}_{1}$
and $\mathbf{C}_{2}$ that are located on its boundary given by \eqref{eq: N-dim boundary},
i.e., they satisfy 
\begin{align}
\det\mathbf{C}_{1} & =\det\mathbb{\mathbf{C}}_{2}=\left(\frac{\hbar}{2}\right)^{2N}\, \label{eq: C1 C2 on boundary}
\end{align}

We recall that covariance matrices are positive definite, $\mathbf{C}_{1},\mathbf{C}_{2}>0$,
and that they must have sufficiently large symplectic eigenvalues
in order to stem from quantum states. Convexity holds if the (positive
definite) convex combination of two covariance matrices, 
\begin{align}
\mathbf{C}(t) & =t\mathbf{C}_{1}+(1-t)\mathbf{C}_{2}\,,\qquad t\in[0,1]\, 
\end{align}
either lies on the boundary of the uncertainty region or in its interior.
This property follows from the fact that the matrix function 
\begin{equation}
g(\mathbf{A})=-\ln\det\mathbf{A}
\end{equation}
is convex \cite{boyd+04}, i.e., the inequality 
\begin{equation}
g(t\mathbf{A}+(1-t)\mathbf{A}^{\prime})\leq tg(\mathbf{A})+(1-t)g(\mathbf{A}^{\prime})\label{eq: ln det convexity}
\end{equation}
holds for any pair of strictly positive definite matrices, $\mathbf{A},\mathbf{A}^{\prime}>0$.
Rewriting \eqref{eq: C1 C2 on boundary} in the form 
\begin{align}
-\ln\det\left(\mathbf{C}_{1}/\hbar\right) & =-\ln\det\left(\mathbf{C}_{2}/\hbar\right)=2N\ln2\, \label{eq: C1 C2 on boundary-1}
\end{align}
one immediately finds that 
\begin{equation}
-\ln\det\left[\left(t\mathbf{C}_{1}+(1-t)\mathbf{C}_{2}\right)/\hbar\right]\leq-t\ln\det\left(\mathbf{C}_{1}/\hbar\right)-(1-t)\ln\det\left(\mathbf{C}_{2}/\hbar\right)=2N\ln2\, 
\end{equation}

Since 
\begin{equation}
\det\left(t\mathbf{C}_{1}+(1-t)\mathbf{C}_{2}\right)\geq\left(\frac{\hbar}{2}\right)^{2N}\,,\quad t\in[0,1]\, \label{eq: convexity for covariance matrices}
\end{equation}
follows, and we have shown that the convex combination of two covariance
matrices on the boundary of the uncertainty region cannot produce
a point outside of it. Equality holds in \eqref{eq: convexity for covariance matrices}
only if $t=0$ or $t=1$. Therefore, states on the boundary cannot
be written as mixtures, which means that the states on the boundary
must be pure states.

Clearly, the argument just given extends to convex combinations of
covariance matrices located \emph{inside} the uncertainty region:
no such combination will produce a covariance matrix on its boundary
or outside of it.

\subsubsection{The Uncertainty Region Has No Pure-State Holes}

We determined the conditions for uncertainty functionals to have extrema
by evaluating them on all \emph{pure} states of $N$ quantum particles.
We now show that the inclusion of mixed states as potential extrema
does not change our findings. It is sufficient to show that all points
of the uncertainty region defined by the inequality \eqref{eq: det form of RS for N particles}
correspond to covariance matrices that stem from pure states.

Recall that any admissible covariance matrix can be diagonalised according
to Williamson's theorem \cite{williamson36,dutta+95} using a suitable
symplectic transformation. Let us order its $N$ finite symplectic
eigenvalues $s_{1}$ to $s_{N}$ from smallest to largest and choose
an integer $M\geq2$ such that $s_{N}\leq M+\nicefrac{1}{2}$ holds.
Suppose now that the $k$-th subsystem resides in the pure state 
\begin{equation}
\ket{\psi_{k}}=\sqrt{t_{k}}\ket{n_{k}=0}+\sqrt{1-t_{k}}\ket{n_{k}=M}\,,\quad k\in\{1,\ldots,N\}\,,\quad t_{k}\in[0,1]\, 
\end{equation}

The variances of position and momentum take the values 
\begin{equation}
\left.\Delta^{2}p_{k}\right|_{\psi_{k}}=\left.\Delta^{2}q_{k}\right|_{\psi_{k}}=(1-t_{k})\left(M+\frac{1}{2}\right)\hbar\,,\quad k\in\{1,\ldots,N\}\,,\quad t_{k}\in[0,1]\, 
\end{equation}
where we use the fact that the expectations of the operators $\hat{p}_{k}$
and $\hat{q}_{k}$ vanish (cf. remark after \scalebox{.95}[1.0]{Equation (\ref{second moments definition})).} Thus, a suitable value of the parameter
$t_{k}$ leads to the desired entries $s_{k}$ on the diagonal of
the covariance matrix, and the covariance of position and momentum
$\hat{p}_{k}$ and $\hat{q}_{k}$ equals zero. \mbox{In addition,} the remaining
off-diagonal matrix elements---associated with the bilinear
operators $\hat{p}_{k}\hat{q}_{k'}$ for $k\neq k'$---also vanish in the product~state 
\begin{equation}
\ket{\Psi}=\ket{\psi_{1}}\otimes\ldots\otimes\ket{\psi_{N}}\, 
\end{equation}

Consequently, there is a pure product state, namely $\ket{\Psi}$,
to generate any desired \emph{diagonal} covariance matrix---which is sufficient to create any admissible \emph{non-diagonal} covariance
matrix, \textls[-15]{simply~by undoing the symplectic transformation used to diagonalize
the initially given covariance~matrix.}

The map from the set of pure states to the interior of the space of
moments is, of course, many-to-one. This can be seen directly by recalling
that each admissible covariance matrix $\mathbf{C}$ can also be obtained
from a Gaussian state characterized by a quadratic form determined
by the matrix $\mathbf{C}$.

\subsubsection{All Moments Arise as Convex Combinations of Two Pure States}

Given any point inside the uncertainty region, one can find infinitely
many convex combinations of two pure Gaussian states on the boundary
that produce the desired $N(2N+1)$ moments. Here~is one way to construct
such pairs. Consider any two-dimensional Euclidean plane that passes
through the origin of the space of moments, $\mathbb{R}^{N(2N+1)}$,
and the given point inside the uncertainty region. The~intersection
of its boundary with the plane is a one-dimensional set of points
that divides the plane into two regions corresponding to acceptable
covariance matrices (forming the uncertainty region) and the rest.
This line inherits convexity from the boundary in the space ${\cal S}$
since any two~points on the curve are, of course, also located on
the high-dimensional boundary.

To conclude the argument, we only need to identify two points on the
boundary such that the line connecting them goes through the point
representing the desired set of moments. It is geometrically obvious
that there exist infinitely many pairs of points on the boundary that
satisfy this requirement. This situation is illustrated in Figure \ref{fig: convex uncertainty region}
in Section \ref{par: convex combinations of pure} for a single continuous
variable where the boundary of the uncertainty region is known to
be a hyperbola.

\subsection{One Continuous Variable: $N=1$\label{sub: One-continuous-variable}}

It is instructive to study the properties of the uncertainty region
for a single continuous variable since the space of moments has only
three dimensions. Even in the absence of entangled states, the uncertainty
region has a number of interesting features as it resembles the Bloch
ball used to visualize the states of a qubit. For one continuous variable,
each point inside the uncertainty region is characterized uniquely
by a triple of numbers, the states on the convex boundary are the
only pure states, and the decomposition of mixed states into pairs
of pure states is clearly not unique. The group of $\mbox{Sp}(2,\mathbb{R})\simeq\mbox{SO}(2,1)$
transformations that leave the uncertainty region invariant play
the role of the $\mbox{SU}(2)$ transformations mapping the Bloch
ball to itself.

We simplify the notation to discuss the case $N=1.$ Renaming the
elements of the $2\times2$ covariance matrix according to 
\begin{equation}
\mathbf{C}=\left(\begin{array}{cc}
\Delta^{2}p & C_{pq}\\
C_{pq} & \Delta^{2}q
\end{array}\right)\equiv\begin{pmatrix}x & w\\
w & y
\end{pmatrix}\, 
\end{equation}
the consistency conditions \eqref{eq: consistency N=00003D00003D1}
take the form

\begin{equation}
xf_{x}=yf_{y}\,,\qquad xf_{w}=-2wf_{y}\, \label{eq: alternative consistency I}
\end{equation}
and 
\begin{equation}
xy-w^{2}=\left(n+\frac{1}{2}\right)^{2}\hbar^{2}\,,\qquad n\in\mathbb{N}_{0}\, \label{eq: alternative consistency RS part}
\end{equation}

The third constraint is \emph{universal }since it does not depend
on the function $f(x,y,w)$ that characterizes an uncertainty functional
$J[\psi]$. It will be convenient to use the variables 
\begin{equation}
u=\frac{1}{2}\left(x+y\right)>0\,,\quad v=\frac{1}{2}\left(x-y\right)\in\mathbb{R}\, 
\end{equation}
to parametrize the points in the \emph{three}-dimensional space of
second moments\emph{,} with coordinates $(u,v,w)^{\top}\in\mathbb{R}^{3}$.
For each non-negative integer, the third condition 
\begin{equation}
u^{2}-v^{2}-w^{2}=e_{n}^{2}\,,\qquad e_{n}=\left(n+\frac{1}{2}\right)\hbar\,,\quad n\in\mathbb{N}_{0}\, \label{eq: uvw hyperboloid}
\end{equation}
determines one sheet of a two-sheeted hyperboloid, located in the
``upper'' half of the space of moments, i.e., $u>0$ and $v,w\in\mathbb{R}$.
The $n$-th sheet---which we call ${\cal E}_{n}$, $n\in\mathbb{N}_{0}$---intersects the $u$-axis at $u=+e_{n}$, and it is in
one-to-one correspondence with the squeezed states originating from
the number state $\ket n$ (cf. \cite{kechrimparis+15}).

The states which satisfy Equation \eqref{eq: uvw hyperboloid} for $n=0$
\emph{saturate} the standard Robertson-Schr\"{o}dinger inequality. Consequently,
not all points in the space of moments can arise as moment triples.
The~accessible part of the space is bounded by the hyperboloid ${\cal E}_{0}$
defined in Equation~\eqref{eq: uvw hyperboloid}, suggesting~us to visualize
the uncertainty region as a solid body with boundary ${\cal E}_{0}$.

We follow the presentation of the multidimensional case in Section \ref{sub: More-than-one},
giving at times alternative proofs of the general results, by appealing
to intuition available in the space of second moments due to its low
dimension.

\subsubsection{The Uncertainty Region Has a Convex Boundary}

Given two mixed quantum states described by density matrices $\hat{\rho}_{1}$
and $\hat{\rho}_{2}$, their convex combinations $\hat{\rho}_{t}=t\hat{\rho}_{1}+(1-t)\hat{\rho}_{2}$,
$t\in[0,1]$, are also quantum states. We now show that the uncertainty
region in the space $\mathbb{R}^{3}$ inherits convexity from the
body of density matrices: any convex combination of the states $\hat{\rho}_{1}$
and $\hat{\rho}_{2}$ with moment triples $\vec{\mu}_{k}=(x_{k},y_{k},w_{k})$,
$k=1,2$, inside the uncertainty region produces another state with
a moment triple also in that region. The boundary of an analogously defined uncertainty region for a quantum spin $s$ \cite{dammeier+15} is not convex. This approach does not use the
convexity of the logarithm of positive definite matrices in \eqref{eq: ln det convexity}.

The moments $x_{k}=\mbox{Tr}(\hat{x}^{2}\hat{\rho}_{k})$, $k=1,2$,
etc., satisfy the Robertson-Schr\"{o}dinger inequality, 
\begin{equation}
x_{k}y_{k}-w_{k}^{2}\geq\frac{\hbar^{2}}{4}\equiv e_{0}^{2}\,,\quad k=1,2\, \label{eq: RS for rho12}
\end{equation}
and the moments of the mixture are given by 
\begin{align}
\sigma_{t} & =t\sigma_{1}+(1-t)\sigma_{2}\,,\qquad\sigma=x,y,w\, 
\end{align}

Writing $\overline{t}=1-t$, the variances of the convex combination
satisfy 
\begin{align}
x_{t}y_{t}-w_{t}^{2} & \geq\left(t^{2}+\overline{t}^{2}\right)e_{0}^{2}+t\overline{t}\left(x_{1}y_{2}+x_{2}y_{1}-2w_{1}w_{2}\right)\,\label{eq: RS for convex combination}
\end{align}
using \eqref{eq: RS for rho12}. Since 
\begin{align*}
x_{1}y_{2}+x_{2}y_{1}-2w_{1}w_{2} & \geq e_{0}^{2}\left(\frac{y_{2}}{y_{1}}+\frac{y_{1}}{y_{2}}\right)+\left(w_{1}\sqrt{\frac{y_{2}}{y_{1}}}-w_{2}\sqrt{\frac{y_{1}}{y_{2}}}\right)^{2}\\
 & \geq2e_{0}^{2}
\end{align*}
holds, the moment triple of the convex combination $\hat{\rho}_{t}$
must also be contained in the uncertainty region,~i.e., 
\begin{equation}
x_{t}y_{t}-w_{t}^{2}\geq\frac{\hbar^{2}}{4}\, \label{eq: RS for rho -final}
\end{equation}

The minimum is obtained only if either $t=0$ or $t=1$, so that the
resulting density matrix must describe a state on the boundary of
the uncertainty region, i.e., a Gaussian state with minimal~uncertainty.

\subsubsection{The Uncertainty Region Has No Pure-State Holes}

Each mixed state $\hat{\rho}$ generates a moment triple $\vec{\mu}$
with components $x=\mbox{Tr}\left(\hat{\rho}\hat{x}^{2}\right)$,
etc., satisfying the Robertson-Schr\"{o}dinger inequality \cite{dodonov+80}.
Thus, the uncertainty region necessarily contains all potential mixed-state
minima $\vec{\mu}$ of a given functional. We want to show that all
moment triples inside the uncertainty region can be obtained through
\emph{pure }states. Two cases occur.

If the triple $\vec{\mu}$ is located on one of the hyperboloids ${\cal E}_{n}$,
$n\in\mathbb{N}_{0}$, then there exists a squeezed number state---i.e., a pure state---which gives rise to the same three
expectations. Hence, the point $\vec{\mu}$ has already been included
in the search for extrema.

Alternatively, the point $\vec{\mu}$ is located between two hyperboloids,
${\cal E}_{n}$ and ${\cal E}_{n+1}$, say, with $n\in\mathbb{N}_{0}$.
Again, there is a pure state with moments given by $\vec{\mu}$. To
see this, we first consider only the line segment with end points
$(u_{n},0,0)$ and $(u_{n+1},0,0)$, which are associated with the
number states $\ket n$ and $\ket{n+1}$, respectively. The moments
of the superposition 
\begin{equation}
\ket n_{t}=\sqrt{t}\ket n+\sqrt{1-t}\ket{n+1},\quad t\in[0,1]\, 
\end{equation}
indeed lead to all moment triples located on the line segment, 
\begin{equation}
\vec{n}_{t}=\left(u_{n+1}+t\left(u_{n}-u_{n+1}\right),0,0\right)\,,\quad t\in[0,1]\, 
\end{equation}
since the matrix elements of the second moments between states of
different parity vanish.

Finally, any moment triple $\vec{\mu}$ off the $u$-axis will lie
on a hyperboloid with a specific value of $t=t_{0}$, for example. This moment
triple can be obtained, however, from the state $\hat{S}(\xi)\ket n_{t_{0}}$,
with a suitable value $\xi$. Using relativistic terminology, the
operator $\hat{S}(\xi$) must induce a Lorentz transformation that
maps the given point on the $u$-axis to the desired point $\vec{\mu}$
on the same hyperboloid.

In conclusion, each triple $\vec{\mu}$ of the uncertainty region
can be obtained from a suitable pure state. Thus, mixed states do
not give rise to candidates for minima different from those associated
with pure~states.

\subsubsection{All Moments Arise as Convex Combinations of Two Pure States \label{par: convex combinations of pure}}

Consider a state $\ket{\xi}$ giving rise to the moment vector $\vec{\xi}=(u_{\xi},v_{\xi},w_{\xi})$
\emph{inside }the uncertainty region. It is possible to identify infinitely
many pairs of Gaussian states on the boundary such that their mixture
reproduces the given triple $\vec{\xi}$.\emph{ }

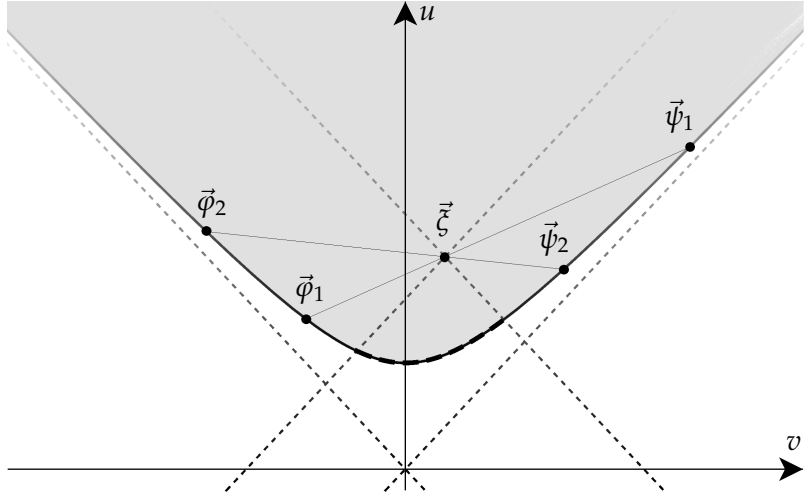
\begin{figure}[h]
	\def\svgwidth{0.7\textwidth}
	\begin{center}
		\begingroup%
		\makeatletter%
		\providecommand\color[2][]{%
			\errmessage{(Inkscape) Color is used for the text in Inkscape, but the package 'color.sty' is not loaded}%
			\renewcommand\color[2][]{}%
		}%
		\providecommand\transparent[1]{%
			\errmessage{(Inkscape) Transparency is used (non-zero) for the text in Inkscape, but the package 'transparent.sty' is not loaded}%
			\renewcommand\transparent[1]{}%
		}%
		\providecommand\rotatebox[2]{#2}%
		\ifx\svgwidth\undefined%
		\setlength{\unitlength}{576bp}%
		\ifx\svgscale\undefined%
		\relax%
		\else%
		\setlength{\unitlength}{\unitlength * \real{\svgscale}}%
		\fi%
		\else%
		\setlength{\unitlength}{\svgwidth}%
		\fi%
		\global\let\svgwidth\undefined%
		\global\let\svgscale\undefined%
		\makeatother%
		\begin{picture}(1,0.61979167)%
		\put(0,0){\includegraphics[width=\unitlength,page=1]{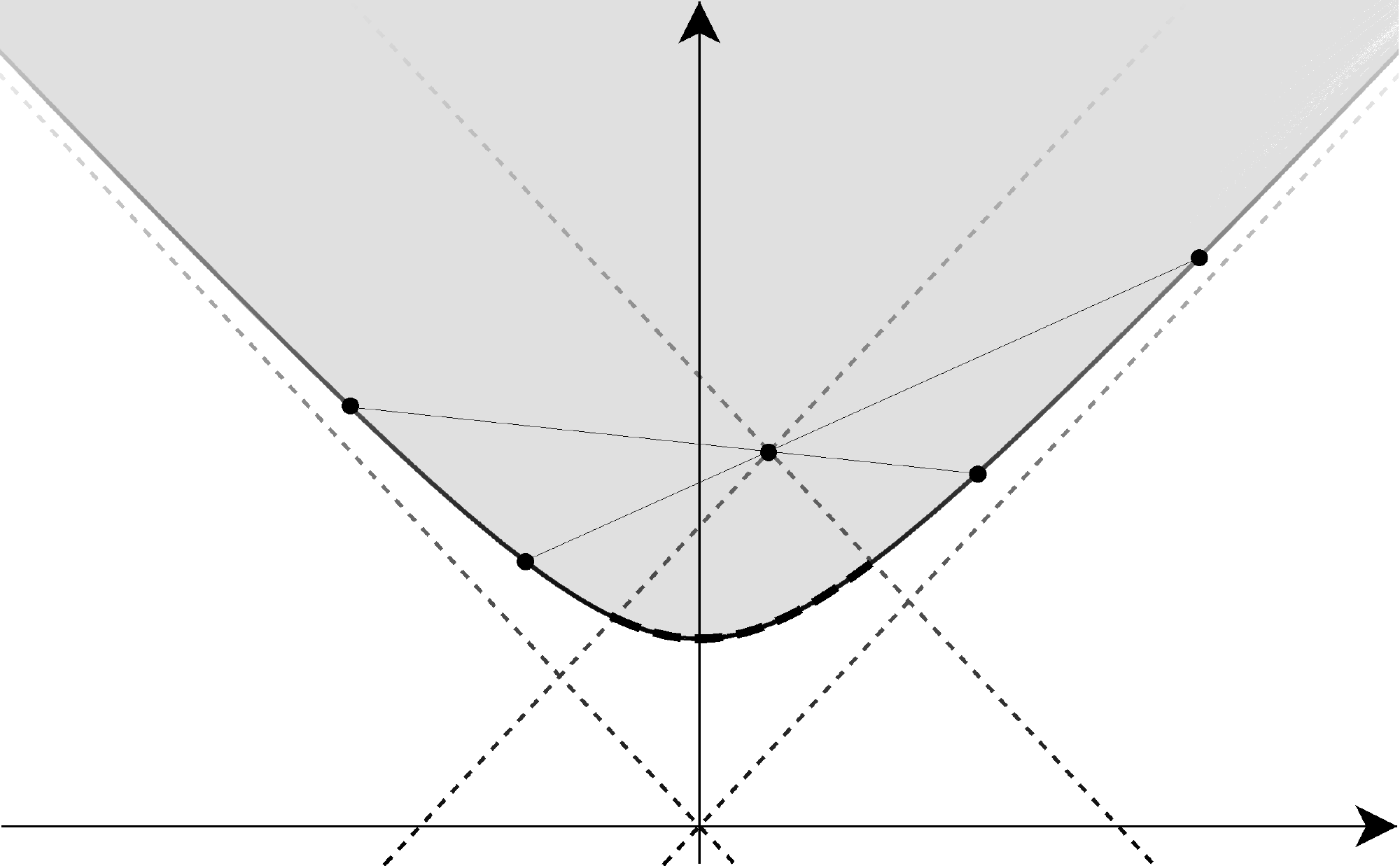}}%
		\put(0.51658604,0.59620661){\color[rgb]{0,0,0}\makebox(0,0)[lb]{\smash{$u$}}}%
		\put(0.97630024,0.05308039){\color[rgb]{0,0,0}\makebox(0,0)[lb]{\smash{$v$}}}%
		\put(0.53743903,0.33175251){\color[rgb]{0,0,0}\makebox(0,0)[lb]{\smash{$\vec{\xi}$}}}%
		\put(0.82369412,0.46350564){\color[rgb]{0,0,0}\makebox(0,0)[lb]{\smash{$\vec{\psi}_1$}}}%
		\put(0.66634867,0.30900378){\color[rgb]{0,0,0}\makebox(0,0)[lb]{\smash{$\vec{\psi}_2$}}}%
		\put(0.36018843,0.24834042){\color[rgb]{0,0,0}\makebox(0,0)[lb]{\smash{$\vec{\varphi}_1$}}}%
		\put(0.23696611,0.35639697){\color[rgb]{0,0,0}\makebox(0,0)[lb]{\smash{$\vec{\varphi}_2$}}}%
		\end{picture}%
		\endgroup%
	\end{center}
	
	\centering{}\protect\caption{\label{fig: convex uncertainty region}Cross-section ($w=0$) of the
		uncertainty region (shaded) illustrating the convexity of its boundary
		$u^{2}-v^{2}-w^{2}=\hbar^{2}/4$; convex combinations of moment triples
		located on the hyperboloid (associated with pure Gaussian states with
		minimal uncertainty) reproduce any given moment vector $\vec{\xi}$
		inside the uncertainty region (the points must be outside of the ``back-ward
		light-cone'' of the point $\vec{\xi}$, indicated by the dashed segment
		of the hyperbola). }
\end{figure}

On the level of moments, it is geometrically obvious that any moment
triple $\vec{\xi}$ can be reached as a~convex combination of two
points located on the boundary (cf. Figure~\ref{fig: convex uncertainty region}).
It is sufficient to consider states with vanishing covariance, $w=0$.
This choice is equivalent to selecting a particular two-dimensional
plane in the space of moments that passes through the origin and
the given moment triple $\vec{\xi}$ (cf. Section \ref{sub: More-than-one}).
Picking any point $\vec{\varphi}$ ``space-like'' relative to $\vec{\xi}$
and located on the hyperboloid, the pair determines a line intersecting
the boundary in a unique point $\vec{\psi}$. Then, the desired point
$\vec{\xi}$ must lie on the line segment $\vec{\xi}(t)=\vec{\varphi}+t(\vec{\psi}-\vec{\varphi})$,
$t\in[0,1]$, connecting the points $\vec{\varphi}$ and $\vec{\psi}$;
it will pass through the point $\vec{\xi}$ if 
\begin{equation}
t_{0}=\frac{u_{\xi}-u_{\varphi}}{u_{\psi}-u_{\varphi}}\equiv\frac{v_{\xi}-v_{\varphi}}{v_{\psi}-v_{\varphi}}\in[0,1]\, 
\end{equation}

When writing the line segment in the form $\vec{\xi}(t)=t\vec{\psi}+(1-t)\vec{\varphi}$,
it becomes obvious that the reasoning valid in the space of moments
extends to quantum states. In other words, the mixture 
\begin{equation}
\hat{\rho}_{t_{0}}=t_{0}\hat{P}_{\psi}+(1-t_{0})\hat{P}_{\varphi}\label{eq: gaussian mixture-1}
\end{equation}
of the rank-1 projectors $\hat{P}_{\psi}=\kb{\psi}{\psi}$ and $\hat{P}_{\varphi}=\kb{\varphi}{\varphi}$
onto Gaussian states on the boundary defines a mixed quantum state
with the desired moment triple $\vec{\xi}$. Clearly, continuously
many other convex combinations of pure states exist that lead to
the same moment triple.

The relationships between quantum states and points inside the uncertainty
region is, of course, many-to-one. For example, the state $\ket 1$
with moment vector $\vec{\xi}_{1}=(9\hbar^{2}/4,0,0)$, i.e., the first
excited state of a harmonic oscillator with unit mass and frequency,
being a pure state, cannot be written as a~mixture of two Gaussian
states. Nevertheless, suitable mixtures of Gaussian states will produce
its moment vector\emph{ }$\vec{\xi}_{1}$. The only moment vectors
$\vec{\xi}$ that cannot be obtained from mixtures are those on the
boundary of the uncertainty region. Here, the relationship between states
and moment vectors is one-to-one, in agreement with the fact that
these Gaussian states are determined uniquely by their covariance
matrix $\mathbf{C}$.

\section{Conclusions}

We have presented a method to systematically determine lower bounds
of uncertainty functionals, defined in terms of second moments of
quantum systems with two or more continuous variables. In~analogy
to the one-dimensional case discussed in \cite{kechrimparis+15},
we find that the states which extremize an~uncertainty functional
of $N$ degrees of freedom must satisfy a (non-standard) eigenvalue
equation that is quadratic in the $2N$ position and momentum operators.
If the quadratic form associated with this operator is positive (or
negative) definite, Williamson's theorem ensures that it can be diagonalised
by a symplectic transformation. In general, the matrix describing
the quadratic form depends on the unknown state suggesting to
solve it in a self-consistent way. The solutions of the resulting
\emph{consistency conditions} determine the set of states that minimise
a given functional. We also introduced the $N(2N+1)$-dimensional
uncertainty region for a system with $N$ continuous variables. We
show that this region is a convex subset of the space of second moments,
and the points located on the boundary correspond to Gaussian states
with minimal uncertainty.

Applying this method to specific functionals associated with quantum
systems described by two~continuous variables, we both re-derived
existing uncertainty relations and previously unknown~ones. We are
not aware of other methods to obtain these inequalities.

One of the new inequalities generalizes an existing inequality that
is capable of detecting entanglement in states of bi-partite particle
systems. This example hints at the possibility to systematically construct
inequalities that can be used for entanglement detection: take an
arbitrary number of EPR-type operators that pairwise commute, and
define a monotonically increasing function of their variances that
is finite at the origin. Typically, the lower bound given by the value
of the functional at the origin will be achieved by an \emph{entangled}
state, and it will be smaller than the value of the functional, which
it can take in any separable state. This bound can be obtained by
solving the consistency conditions \eqref{CCsProd} for product states
as described in Section \ref{sec:NoCorrelations}. Clearly, a violation
of the pure-state bound will detect the presence of an entangled state.
The details of this construction will be left to a future publication.


\subsection*{Acknowledgements}
S.K. has been supported via the act ``Scholarship Programme of the State Scholarship Foundation (IKY)
by the procedure of individual assessment, of 2011--2012''
by resources of the Operational Programme for Education and Lifelong
Learning of the ESF and of the NSF, 2007\textendash 2013, as well
as by the \mbox{WW Smith Fund.}




\end{document}